\documentclass{article}

\usepackage{listings}
\usepackage{hyperref}

\title{Monads and ``do" notation in the Wolfram Language}
\author{Kacper Topolnicki}
\date{\today}

\begin{document}

\maketitle

\begin{abstract}
This paper describes a categorical interpretation of the
\emph{Wolfram Language}
and introduces a simple implementation of monadic types and the ``do" notation. 
The monadic style of programming combined with the many built 
in functions of the \emph{Wolfram Language} has potential to be a 
powerful tool in writing \emph{Wolfram Language} code. Additionally,
using pure functions and the ``do" notation can result in programs that
are very predictable and easy to parallelize.

\end{abstract}

\section{Introduction}

The motivation for this work started from a
practical problem when the author was
writing a \emph{Wolfram Language} (WL) application that aimed to abstract away
the notion of scalar valued functions for numerical computations.
The aim was for the user to specify, in the WL, the number of arguments for 
a function, which arguments are discrete indices and which arguments
are floating point numbers, for floating point arguments - the points
and weights used for interpolations and integration over these arguments and
finally specify which arguments can be used for purposes of code parallelization.
Given this input the goal was for the application to create a \emph{FORTRAN 90}
module and a directory with supplementary data. 
The user could then import this module in his or her program and work with
an abstraction of the scalar valued function: store function values on 
different cores and handle them using different \emph{MPI} and \emph{OPENMP}
threads, interpolate function values and integrate over function arguments.

The application requires further testing but is feature complete.
Achieving this was possible, to a large extent, due to a choice to base
the implementation around monadic types. This combined with the many 
powerful built in functions in the WL resulted in a very processing tool.
The monadic style of programming can also be appropriate for other problems and
in this paper a simple approach to include monads and the ``do" notation 
in the WL is shown. This work was also influenced by 
an book by Bartosz Milewski \cite{bmilew} containing
an excellent introduction to Category Therory. 

The text is organized as follows. Section \ref{categorical_interpretation}
introduces a categorical interpretation of the WL. Section
\ref{monadic_types}
describes the implementation of monadic types and the ``do" notation. Section
\ref{examples}
contains an examples and tips for practical implementations. Finally
section \ref{summary} contains a summary.
The notation used in WL pseudocode shown in some parts of the text uses
comments \verb|(*...*)| to hold additional information intended for the reader.

\section{Categorical interpretation of the \emph{Wolfram Language}}
\label{categorical_interpretation}

In order to think about the WL as a category a subset of the WL will be considered consisting
of functions defined in the following manner:
\begin{lstlisting}
f[(*pattern a*)] := (*expression*)
\end{lstlisting}
where \emph{pattern a} is a WL language pattern expression that specifies the
arguments accepted by the function \emph{f} and \emph{expression} is a WL
expression that can be constructed from the named parts of \emph{pattern a}.
This is not directly enforced by the WL but ideally all possible results 
of \emph{f} match a different pattern, \emph{pattern b}, so that:
\begin{lstlisting}
MatchQ[f[(*expression matching pattern a*)],(*pattern b*)]
\end{lstlisting}
is always true. In some places Curried functions will be used and this can be
interpreted as replacing \verb|f[x]|, \verb|f[x][y]|, \ldots for \verb|f|
in the pseudo code above and treating the replacement as a whole set of
functions built from the different values of \verb|x|, \verb|y|, \ldots

It is seems natural to treat this subset of the WL as a category with morphisms
being functions (e.g. \verb|f|, \verb|f[x]|, \verb|f[x][y]|, \ldots) and objects
being types defined by patterns (e.g. the set of all WL expressions that match \emph{pattern a} 
defines type \emph{a} and the set of expressions that match \emph{pattern b}
defines type \emph{b}). In this context a monad is a type function that takes a
pattern that defines one type and returns a new pattern that defines another
type.

\section{Introduction of monadic types and the ``do" notation}
\label{monadic_types}

A user can implement a monad \emph{m} in the WL by supplying three definitions.
The first definition:
\begin{lstlisting}
pattern[m] = (*pattern m a*)
\end{lstlisting}
provides a pattern that matches a WL expression if it is an \emph{m} monad for any type argument
\emph{a}. The second definition implements the monadic \emph{return} function:
\begin{lstlisting}
return[m][x_] := (*expression matching pattern[m]*)
\end{lstlisting}
if the type of \emph{x} is \emph{a} then this expression should be a
\emph{m} monad for type argument \emph{a}. In particular the resulting
WL expression should match \verb|pattern[m]|. The final definition is 
an implementation of the monadic
\emph{bind} operator for \emph{m}:
\begin{lstlisting}
bind[m][ma_ , aTOmb_] := (*implementation of bind*)
\end{lstlisting}
where \emph{ma} is an \emph{m} monad for type argument \emph{a} and
\emph{aTOmb} is a function taking an element of type \emph{a} to an 
\emph{m} monad for type argument \emph{m}. The user does not have
to supply additional pattern constraints in \verb|bind|, however 
it is up to the user to make sure that these definitions result in
an object that satisfies the monad laws.

Having these three definitions for monad \emph{m}, the implementation of the
``do" notation in the WL is straightforward and consists of two parts. 
The first is a recursive definition that reduces the \verb|do[m]| expression
by applying the user defined \verb|bind[m]| operator:
\begin{lstlisting}
do[m_][x___ , y_ , r:Except[_LeftArrow]]:= 
   If[MatchQ[y , LeftArrow[_ , _]] , 
      do[m][x , 
         chk[m][
         bnd[m][snd[y] , 
            Function[Evaluate[fst[y]] , r]]]
         ],
      do[m][x , 
         chk[m][bnd[m][y , 
            Function[Unique["nonExistantVariable"] , r]]]
      ]
   ];
\end{lstlisting}
Here \verb|bnd[m]| will be replaced by the appropriate user supplied
\verb|bind[m]|
function when the first argument matches \verb|pattern[m]| and the second
argument is a function, \verb|fst|, \verb|snd| are helper functions that take the left and
right expression from a \verb|LeftArrow| statement, and finally \verb|chk[m]|
is a function that checks the result of \verb|bind[m]| against the user defined 
\verb|pattern[m]|. Once \verb|do[m]| is reduced to a contain a single expression 
the final expression is checked and returned:
\begin{lstlisting}
do[m_][mx_]:=chk[m][mx];
\end{lstlisting}
These definitions together with the supplementary functions are provided
for the readers convenience at \cite{repos}. This repository contains a 
small WL package and a number of examples. 

\section{Hanoi Tower example}
\label{examples}

This section takes a look  
at the classic Hanoi Tower puzzle example solved in \cite{repos}.
The puzzle
consists of three poles onto which stacks of disks with different sizes $1 , 2 , 3 , \ldots$
can be placed. At the beginning of the puzzle all disks are stacked on the first pole
with smaller disks being placed on top of larger ones. The aim of the puzzle is to 
transport all disks from the first pole to the third pole in discrete steps. In each 
step only one disk can be taken from the top of a stack and placed on a larger disk
or on an empty pole.

The puzzle is solved in the \emph{hanoi\_tower\_example.nb} notebook from \cite{repos}.
The notebook starts a definition 
of a pattern for the Hanoi Tower puzzle:
\begin{lstlisting}
towers = {_List , _List , _List};
\end{lstlisting}
The puzzle will be represented by a list containing three elements. Each 
element is a list and corresponds to a single pole of the puzzle.
Elements of this list will be numbers corresponding to disc sizes. 
Another pattern is provided for expressions that will be used to record
the moves that lead to the solution:
\begin{lstlisting}
move = {_ -> _ , {_List , _List , _List}};
\end{lstlisting}
Moves will be two element lists, the first element (e.g. \verb|1->2|)
will record the number of the pole from which the disk was taken and the number
of the pole onto which the disk was dropped (e.g. \verb|1| , \verb|2|). The
second element will contain the current configuration of the puzzle.

Next three definitions are provided for the \emph{hT} monad that will be used in
a recursive solution to the puzzle. The first definition is a pattern that will match any
WL expression that is an \emph{hT} monad for any type argument:
\begin{lstlisting}
pattern[hT] = hT[towers , {move ...}];
\end{lstlisting}
Next the return function is defined as (please note that an empty list also matches \verb|{move ...}|):
\begin{lstlisting}
return[hT][x_] := hT[x , {}];
\end{lstlisting}
Finaly the bind
operator for \emph{hT} is given (please note that no additional pattern
constraints are necessary for \verb|ma| and \verb|aTomb|):
\begin{lstlisting}
bind[hT][ma_ , aTomb_] := 
   With[
      {val = aTomb[grab[ma]]}, 
      hT[grab[val] , join[rest[ma], rest[val]]]
   ];
\end{lstlisting}
where \verb|grab| and \verb|rest| are helper functions:
\begin{lstlisting}
grab[ma : pattern[hT]] := ma[[1]];
rest[ma : pattern[hT]] := ma[[2]];
\end{lstlisting}
and \verb|join| provides a lazy version of the \verb|Join| function:
\begin{lstlisting}
join[a : {move ...} , b : {move ...}] := Join[a , b];
\end{lstlisting}

Only one action is provided for this monad with two definitions. 
The first one moves a single disk
of puzzle \verb|tower| from pole number \verb|from| to pole number \verb|to|:
\begin{lstlisting}
moveDiscs[from_ , to_ , 1][tower : towers] :=
   Module[{newtower},
      newtower = ReplacePart[
         ReplacePart[
            tower , 
            to -> Join[tower[[from]][[1 ;; 1]], tower[[to]]]
         ] , 
         from -> tower[[from , 2 ;;]]
      ];
      hT[newtower , {{from -> to , newtower}}]
   ];
\end{lstlisting}
The second one moves \verb|n| disks from pole number \verb|from|
to pole number \verb|to|. This time the definition uses the ``do" notation and recursion:
\begin{lstlisting}
moveDiscs[from_ , to_ , n_][tower : towers] :=
   do[hT][
      toOther <- moveDiscs[
                  from , 
                  other[from , to] , 
                  n - 1][tower],
      toGoal <- moveDiscs[
                  from , 
                  to , 
                  1][toOther],
      finalMove <- moveDiscs[
                     other[from , to] , 
                     to , 
                     n - 1][toGoal],
      return[hT][finalMove]
   ];
\end{lstlisting}
where \verb|<-| is the \verb|LeftArrow| operator entered in the WL
using \verb|<ESC><-<ESC>| and \verb|other[i , j]| returns a pole number
that is different then \verb|i| and \verb|j|. The procedure
of solving the puzzle can be read off directly from the code above.
First $n-1$ disks are moved from pole number \verb|from| to a pole that is
different from \verb|from| and \verb|to|. Next, one disk is moved from pole
number \verb|from| to pole number \verb|to|. Finaly the $n-1$ disks are moved
from the other pole to the \verb|to| pole.

A solution to the puzzle with three disks on the first pole in the initial
configuration
can be obtained by evaluating:
\begin{lstlisting}
moveDiscs[1 , 3 , 3][makeTower[3]]
\end{lstlisting}
where \verb|makeTower[3]| prepares the initial configuration of the puzzle.
The result is:
\begin{lstlisting}
hT[
   {{},{},{1,2,3}},
   {
      {1->3,{{2,3},{},{1}}},
      {1->2,{{3},{2},{1}}},
      {3->2,{{3},{1,2},{}}},
      {1->3,{{},{1,2},{3}}},
      {2->1,{{1},{2},{3}}},
      {2->3,{{1},{},{2,3}}},
      {1->3,{{},{},{1,2,3}}}
   }
]
\end{lstlisting}
The first element of this expression is the final configuration of the solved
puzzle. The second element is a list containing the moves used to obtain the
solution. 

The repository \cite{repos} contains a couple more simple examples including a
\emph{list} monad and a \emph{maybe} monad. The reader is encouraged to download
and explore these notebooks.

\section{Summary}
\label{summary}

The \emph{Wolfram Language} comes with a very powerful set of built in
functions. The flexibility of this language makes it possible to program
in many different styles. Such freedom, however, can be overwhelming and 
it is a good idea to conform to styles of programming that are tailored
to particular problems. Combining the monadic style of programming with
the powerful standard library of the \emph{Wolfram Language} can be very
useful for problems that require a functional solution.

\end{document}